\documentstyle[prl,aps,multicol,epsf]{revtex}
\renewcommand{\narrowtext}{\begin{multicols}{2} \global\columnwidth20.5pc}
\renewcommand{\widetext}{\end{multicols} \global\columnwidth42.5pc} 

\begin{document}

\newcommand{\be}{\begin{equation}}
\newcommand{\ee}{\end{equation}}
\newcommand{\bea}{\begin{eqnarray}}
\newcommand{\eea}{\end{eqnarray}}
\newcommand{\nt}{\narrowtext}
\newcommand{\wt}{\widetext}

\title{Interacting random Dirac fermions in superconducting cuprates} 
\author{D.V. Khveshchenko$^{1}$, A.G. Yashenkin$^{2,4,\dagger}$, 
and I.V.Gornyi$^{3,4,{\dagger\dagger}}$}
\address{$^1$ Department of Physics and Astronomy, 
University of North Carolina, Chapel Hill, NC 27599\\
$^2$ Department of Physics, University of Florida, Gainesville, FL 32611\\
$^3$ Institut fur Nanotechnologie, Forschungszentrum Karlsruhe,
76021 Karlsruhe, Germany\\
$^4$ NORDITA, Blegdamsvej 17, Copenhagen 2100, Denmark}
\maketitle

\begin{abstract}
We study the effects of quasiparticle interactions on disorder-induced
localization of Dirac-like nodal excitations in superconducting high-$T_c$ cuprates. 
As suggested by the experimental ARPES and terahertz conductivity 
data in $Bi_2Sr_2CaCu_2O_{8+\delta}$, we focus on
the interactions mediated by the order parameter fluctuations  
near an incipient second pairing transition $d\to d+is$.
We find interaction corrections to the density of states,
specific heat, and conductivity as well as phase and energy relaxation rates
and assess the applicability of the recent localization scenarios for
non-interacting random Dirac fermions to the cuprates. 
\end{abstract}
\nt
In the past few years, the problem of disordered two-dimensional (2D) Dirac 
fermions received much of attention, 
as it provides an effective description for the random bond Ising model, 
network models of Quantum Hall plateau transitions, and some other statistical 
problems. 
Also, dynamical Dirac fermions can be used to conveniently 
describe low-energy excitations in a variety of correlated systems, such as
p-wave (e.g., $Ru_2SrO_4$) and layered  
d-wave (high-$T_c$ cuprates) superconductors 
and superfluids ($He^3$), and zero-gap semiconductors (e.g., graphene sheets).
 
The Hamiltonian of a generic disordered gapless superconductor  
and, in particular, planar $d$-wave system,   
possesses an additional discrete symmetry of charge conjugation
which gives rise to as many as
seven novel random Gaussian ensembles corresponding to 
different patterns of spin rotational and time
reversal symmetry breaking \cite{Altland}.
Moreover, in a stark contrast with the conventional case of a normal metal  
with extended Fermi surface the density of states
(DOS) of the non-interacting Dirac fermions in
a 2D d-wave superconductor is strongly affected by disorder \cite{Lee}.
Furthermore, depending on the concrete model for disorder, such as 
isotropic versus predominantly forward potential impurity scattering,  
the DOS of the random Dirac fermions can exhibit different low-energy
asymptotic behaviors even within the same random ensemble 
\cite{Tsvelik,Senthil1,Senthil2,Pepin,Zirnbauer,Mudry}.

The multitude of different regimes and crossovers predicted for the 
non-interacting random Dirac fermions 
raises a question about their observability in 
such realistic $d$-wave systems as the high-$T_c$ superconductors 
where quasiparticle interactions are believed to be important.
Moreover, the localization theory still remains incomplete
without an extra input in the form of quasiparticle dephasing
rate which controls the magnitude of disorder-induced localization
corrections in the infinite system. 

Thus far, despite the continuing progress in understaning 
of the non-interacting (de)localization phenomena in the $d$-wave systems, 
the above issues did not receive enough attention. In the present paper, 
we fill in this gap by investigating the effects of physically relevant 
quasiparticle interactions on the localization properties
of the Dirac-like nodal excitations in the superconducting cuprates.

In the Nambu spinor representation, the quasiparticle 
(retarded) Green function reads as  
\be
\hat{G}^R_{\bf k}(\epsilon)=[(\epsilon - \Sigma^R_{\bf k}(\epsilon))\hat{\tau_0}
+\xi_{\bf k}\hat{\tau_3}+\Delta_{\bf k}\hat{\tau_1}]^{-1}
\label{green}		    
\ee
where we introduced the $2\times 2$ unity matrix ${\hat\tau}_0$ 
in addition to the triplet of the Pauli matrices 
$\hat\tau_{1,2,3}$.

The bare quasiparticle spectrum 
which is composed of the normal state dispersion $\xi_{\bf k}$ and the
$d_{x^2-y^2}$-symmetrical gap $\Delta_{\bf k}$ becomes 
linear  $E_{\bf k}=[ \xi_{\bf k}^2 + \Delta_{\bf k}^2 ]^{1/2}
\approx [( {\bf v}_{f} {\delta\bf k})^2 + ( {\bf v}_{g} {\delta\bf k})^2 ]^{1/2}$  
for momenta near each of the four gap nodes located at 
$\pm {\bf K}_{1,2}=\pm(\pm k_F, k_F)/{\sqrt 2}$. 
The pairs of orthogonal vectors  
${\bf v}_{f} = (\partial \xi_{\bf k}/\partial {\bf k})$ and
${\bf v}_{g} = (\partial \Delta_{\bf k}/\partial {\bf k})$ 
correspond to the components of the quasiparticle group 
velocity which are normal
and tangent to the fudicial Fermi surface, respectively.

By using the standard DOS definition
\be
\nu(\epsilon)=-\frac{1}{\pi}{\rm Tr}\sum_{\bf k}{\rm Im}G^R_{\bf k}(\epsilon)
\label{DOS} 
\ee
and linearizing the spectrum near the nodes,
one readily recovers the linear 
DOS $\nu(\epsilon)=|\epsilon|/(\pi v_fv_g)$ of the nodal low-energy excitations in a 
clean 2D d-wave superconductor.

The origin of the quasiparticle interactions in the cuprates remains a subject
of an ongoing debate. 
The recent  ARPES \cite{Valla} and optical (terahertz) conductivity
\cite{Corson} data in $Bi_2Sr_2CaCu_2O_{8+\delta}$
seem to rule out any short-ranged coupling that would have resulted in
a $T^3$ temperature dependence of the 
inverse inelastic quasiparticle lifetime \cite{Scalapino}.
Instead, these data show an approximately linear behavior  
which is suggestive of the possibility of 
some sort of a quantum-critical behavior \cite{Sachdev}.

In the common scenario of a quantum-critical point (QCP), 
the long-ranged interactions between the quasiparticles 
result from their exchange by fluctuations of the new order parameter.
The recent analysis of different incipient orderings   
singled out one of the second pairing transitions $d_{x^2-y^2}\to d_{x^2-y^2}+is
(id^\prime_{xy})$ as the most plausible candidate \cite{Sachdev}. 
Alternatively, one can also consider the long-ranged interaction
 due to charge fluctuations in the vicinity 
of a crystal structural instability \cite{DiCastro}.
Yet another type of interaction mediated by the   
exchange of antiferromagnetic fluctuations with momenta
${\bf q}\approx{\bf Q}$ where ${\bf Q}=(\pm\pi,\pi)$  
has been studied extensively  
in the context of the normal state of the cuprates and recently
extended to the superconducting regime \cite{Chubukov}, 
albeit with no disorder included.

In order to put all
of the above quasiparticle interactions into one unifying framework
we consider a generic form of the bosonic collective mode propagator 
\be
\hat{V}^R_i(\omega, {\bf q})=
\frac{V_{i}{\hat{\tau_i}}\otimes{\hat{\tau_i}}}
{(\omega/c)^2 - ({\bf Q}-{\bf q})^2 -\lambda^2- V_{i}\Pi^R_i(\omega,{\bf q})}
\label{interaction}
\ee
which describes the interactions in the 
spin, pairing, and charge channels for $i=0,2$, and $3$, respectively.
In Eq.(\ref{interaction}) the parameter $\lambda^2$ controls the  
proximity to the QCP, and 
the fermion polarization functions $\Pi^R_i(\omega,{\bf q})$
account for the feedback effect of the nodal Dirac fermions on the spectrum of
the collective mode.

As regards modeling the disorder, we consider the experimentally relevant 
case of time-reversal and spin-rotational invariant scattering due to 
non-magnetic impurities (class 
CI in the classification chartered in Ref.\cite{Altland}) which 
we treat as 
scatterers with density $n_i$ coupled to the nodal Dirac fermions via 
the vertex $U_{{\bf k},{\bf k}^\prime}\hat\tau_3$. The latter 
accounts for arbirary strength and momentum dependence
of disorder scattering and gives 
rise to the scattering $T$-matrix 
\be
{\hat T}_{{\bf k},{\bf k}^\prime}(\epsilon)=
\sum_{{\bf k}^{\prime\prime}}
[\delta_{{\bf k},{\bf k}^{\prime\prime}}-U_{{\bf k},{\bf k}^{\prime\prime}}
{\hat\tau_3}{\hat G}^R_{{\bf k}^{\prime\prime}}(\epsilon)]^{-1}
U_{{\bf k}^{\prime\prime},{\bf k}^{\prime}} {\hat\tau_3}
\label{Tmatrix}
\ee
where the first and second terms in the brackets 
correspond to the Gaussian and Unitary 
limits, respectively \cite{Lee}. In the earlier studies employing 
non-abelian bosonization and nonlinear
$\sigma$-model \cite{Tsvelik,Senthil1,Senthil2,Zirnbauer} the latter regime 
describing strong impurities has 
remained unattainable, and a markedly different behavior
predicted in this case \cite{Pepin,Mudry} could not be readily addressed. 

The first insight into the random Dirac problem 
can be obtained from the self-consistent equation 
derived in the standard non-crossed diagrammatic approximation \cite{Lee} 
\be
\hat{\Sigma}^R_{\bf k}(\epsilon)=
{n_i\sum_{{\bf k}^\prime}}
{\hat T}_{{\bf k},{\bf k}^\prime}(\epsilon)
{\hat G}^R_{{\bf k}^\prime}(\epsilon)
{\hat T}_{{\bf k}^\prime,{\bf k}}(-\epsilon)
\label{sigma}
\ee
As shown below, our neglecting the crossed diagrams 
is well justified by the large anisotropy of the Dirac spectrum $(v_F/v_g\sim 20)$ observed 
in the cuprates \cite{Chiao}. 

Provided that $n_i|{\hat T}_{{\bf k},{\bf k}^\prime}(\epsilon)|/v_fv_g<<1$, 
for all the relevant energies and momenta,  
the solution of Eq.(\ref{sigma}) manifests the emergence of a new energy scale
$-Im\Sigma^R(0)=\gamma<<\Delta$  which separates between the ballistic and diffusive 
regimes and gives rise to the finite DOS 
\be
\nu_0=\frac{2\gamma}{\pi^2 v_f v_g}\ln\frac{\Delta}{\gamma}
\label{bareDOS}
\ee
In the non-crossed approximation, the spin and thermal conductivities 
which obey the Wiedemann-Franz law appear to be independent of disorder  
($\hbar = k_B=1$):
\be
\sigma_s={3\kappa\over 4\pi^2T}=
\frac{1}{4\pi^2}\frac{v_f^2+v_g^2}{v_f v_g},  
\label{baresigma}
\ee
while the charge conductivity
receives non-universal vertex corrections \cite{Lee}. This 
is consistent with the fact that, unlike the charge of quasiparticles,
their spin and energy are conserved, and, therefore,
in the absence of localization they can both propagate diffusively 
\cite{Senthil1}. 

In the diffusive ($\epsilon<\gamma$) regime, the mean field DOS 
(\ref{bareDOS})  
and the "universal" conductivities (\ref{baresigma}) alike become subject to 
further corrections which stem from both disorder-induced weak localization and 
interference between disorder scattering and quasiparticle interactions.
In order to compute these corrections one needs explicit
expressions for all the gapless diffusion (${\cal D}$) and Cooperon 
(${\cal C}$) modes 
of the random Dirac problem beyond the ergodic limit studied in Ref.\cite{Altland}. 

The abovementioned strong anisotropy of the Dirac spectrum in the cuprates
implies $\sigma_s\gg 1$ which,
apart from justifying the use of Eq.5 for calculating the fermion self-energy, 
allows one to resort to the standard ladder approximation for 
the propagators $\hat{\cal D} (\hat{\cal C})=\sum_{ij} {\cal D}_{ij} ({\cal C}_{ij}) 
{\hat\tau}_i\otimes{\hat \tau}_{j}$
expanded in the basis of the tensor products of the ${\hat \tau}_i$-matrices.

By analogy with a normal metal, we first consider the ladders formed by one
retarded and another advanced Green functions (RA-ladder)
and find the following singular contributions to the propagators of the soft modes  
\be
{\cal D}_{ij} ({\cal C}_{ij})
(\epsilon, \epsilon^\prime, {\bf q}) = \delta_{ij} 
\frac{\gamma^2}{\pi\nu_{0}} 
\frac{\eta^{D(C)}_{i}}{Dq^2 - i (\epsilon-\epsilon^\prime)}
\label{diffuson}
\ee
with the amplitudes $\eta^{D}_i=(1,1,1,1)$ and $\eta^{C}_i=(1,1,-1,1)$ 
corresponding to the $\cal D$ and $\cal C$ propagators, respectively 
\cite{us}.

As previously pointed out \cite{Senthil1},
the diffusion coefficient $D$ appearing in Eq.(\ref{diffuson}) must satisfy  
the Einstein relation $\sigma_{s}= {D} \nu_0/4$.
In our approach, this happens naturally, once 
the real part of the fermion self-energy ${\rm Re}{\hat\Sigma}
^R_{\bf k}(\omega)$ from Eq.(\ref{sigma}) is accounted for in
the ladder equations. 

As opposed to the case
of a normal metal, in a superconductor the gapless
poles also appear in the RR (AA)-ladders
due to the combination of processes of impurity scattering and  
Andreev reflection \cite{Altland}.
The corresponding propagators ${\overline{\cal D}}({\overline{\cal C}})$ 
are related to (\ref{diffuson}) by 
virtue of the charge conjugation 
performed on one of the two lines of the ladder:
\be
{\overline{\cal D}} ({{\overline{\cal C}}})
(\epsilon,\epsilon^\prime, {\bf q}) = - \sum_i {\cal C}_{ij} ({\cal D}_{ij})
(\epsilon, -\epsilon^\prime, {\bf q})(\hat\tau_2\hat\tau^*_i\hat\tau_2)
\otimes\hat\tau_i 
\label{anomalousdiffuson}
\ee
which can be cast in the form of Eq.(\ref{diffuson}) 
with ${\overline \eta}^{D}_i=(-1,1,-1,1)$ and 
${\overline \eta}^{C}_i=(-1,1,1,1)$.

A straightforward analysis 
shows that it is the RR-Cooperon $\overline{\cal C}$ which appears to be
solely responsible for the first order weak localization DOS correction \cite{Senthil2} 
\be
\delta_{wl}\nu(\epsilon)= -{1\over \pi^2D}\ln{\gamma\over \epsilon}
\label{wlDOS}
\ee
Eq.(\ref{wlDOS}) indicates that perturbation 
expansion remains well under control for energies above the
characteristic scale $\sim\gamma\exp(-4\pi^2\sigma_s)$ where it 
eventually breaks down.

Besides the above soft modes, in the case of a nearly nested (square-like) 
normal state Fermi surface there might also exist additional (pseudo)Goldstone ones 
with gaps of the order of chemical potential $\mu$ of the bare electrons. 
In the $d$-wave superconducting state, the corresponding poles 
occur in both RA- and RR(AA)-ladders 
when the sum of the two momenta carried by the lines equals $\bf Q$, 
provided that the normal state dispersion satisfies 
$\xi_{\bf k}\approx -\xi_{{\bf k}\pm {\bf Q}}$. 
In the Unitary limit, the presence of 
these extra modes reverses the sign of the localization correction (\ref{wlDOS}) 
in the energy interval $\mu^2/\gamma < \epsilon < \gamma$ \cite{us}, 
resulting in a behavior which is reminiscent of that predicted in Refs.\cite{Pepin,Mudry}
for $\mu=0$. Nonetheless, at yet lower energies 
the DOS will eventually get suppressed, in accord with the 
scenario of Refs.\cite{Senthil1,Senthil2,Zirnbauer}.

Next, we calculate the 
interaction (Altshuler-Aronov type) corrections which stem from the  
interplay between disorder and quasiparticle interactions (\ref{interaction}).
As usual, the latter can be divided
into the exchange and Hartree contributions.

Upon computing the diffusion-dressed interaction
vertices, we find that away from half-filling 
$(2{\bf K}_{1,2}\neq{\bf Q})$ no diffusion poles occur 
for any finite transferred momenta ${\bf Q}\neq 0$
which limits the subsequent discussion to the intra-node inelastic scattering.

Furthermore, the vertices $\hat\tau_{1,3}$
undergo no singular diffusion dressing, as they 
correspond to the two spatial components of the
Dirac fermion current operator.
By contrast, both vertices $\hat\tau_{0}$ and $\hat\tau_2$ 
which represent the Dirac fermion density and
mass operators do develop diffusion 
poles in the RA- and RR(AA)-channels, respectively. 

In light of the above, from now on we focus on
the coupling ${\hat V}^R_2(\omega,{\bf q})$ mediated
by the fluctuations of a secondary pairing order parameter with ${\bf Q}=0$
which was suggested as a possible scattering mechanism in the high temperature
(ballistic) regime \cite{Sachdev}. Here we restrict our attention to the 
case of the $is$ secondary pairing, for the case of the
$id_{xy}$ pairing appears to be somewhat more intricate \cite{newAltland}.
Unlike any other before mentioned types of interactions, it 
is strongly enhanced by disorder, thus producing the diffusive DOS correction
\vspace{1.0cm}   
$$
{\delta_{ex}\nu(\epsilon)\over \nu_0}=
\hspace{8.0cm}
$$
\bea
\int^{\infty}_{-\infty}
 {d\omega\over 2\pi}
 \sum_{\bf q}
\left[-{1\over 2}\tanh{\epsilon-\omega\over 2T}
Im { {V^R_2(\omega,{\bf q})} \over {(Dq^2-i(2\epsilon-\omega))^2} }
\right.
\cr
\left.
+\coth{\omega\over 2T}
Im V^R_2(\omega,{\bf q}) Re {1\over (Dq^2-i(2\epsilon-\omega))^2}
\right]
\label{exDOS}
\eea
In the conserving approximation, the above vertex 
renormalization is necessarily
accompanied by the singular polarization of the disordered Dirac fermions
\be
\Pi^R_{2}(\omega,{\bf q})=
\nu_0\ln {\gamma\over Dq^2-i\omega}
\label{polarization}
\ee  
Close to the QCP ($\lambda^2<V_{2}\nu_0$),  
the term (\ref{polarization}) dominates
over the other terms in the denominator of Eq.(\ref{interaction}), thereby resulting
in the intrinsically attractive coupling ${\hat V}^R_2(\omega,{\bf q})
\approx -1/\Pi^R_2(\omega,{\bf q})$ which yields a positive DOS correction 
\be
\delta_{ex}\nu(\epsilon)= {1\over 4\pi^2D}\ln|\ln{\gamma\over \epsilon}|
\label{lnlnDOS}
\ee
Thus, in contrast to the exchange DOS correction  
in a normal metal, in our case $\delta\nu_{ex}$ 
appears to be substantially weaker than the effect of weak localization
(\ref{wlDOS}). 

Farther away from the QCP ($\lambda^2>V_{2}\nu_0$), the bosonic propagator 
(\ref{interaction}) becomes effectively short-ranged, and 
the (positive) interaction correction to DOS becomes explicitly dependent   
upon the coupling strength: 
$\delta_{ex}\nu(\epsilon)\sim (V_{2}\nu_0/\lambda^2)\ln\gamma/\epsilon$,
while the corresponding Hartree term vanishes identically.

The latter regime appears to be partly similar to the case of a short-ranged 
non-singular ferromagnetic coupling  of the local quasiparticle 
spin densities in a gapless bulk superconductor 
which yields the interaction DOS correction 
of the same functional form and sign as the localization one \cite{Fisher}. 

The DOS correction (\ref{lnlnDOS}) can also be deduced from the 
singular contributions to thermodynamic quantities, e.g., specific heat
\bea
\delta C(T)= T{d^2\over dT^2}
\int^{\infty}_0{d\omega\over 2\pi}[\coth({\omega\over 2T})+1]\nonumber\\
\sum_{\bf q}tan^{-1}
{Im\Pi^R_2(\omega,{\bf q})\over Re\Pi^R_2(\omega,q)}
={T\over 2\pi^2D} \ln |\ln{\gamma\over T}| 
\label{specific heat}
\eea
calculated for $\lambda\to 0$. 
Accordingly, the exchange contribution 
to the conductivity correction is given by the expression 
\bea
\delta_{ex}\sigma_s(T)=-2\sigma_s Im \int^{\infty}_{-\infty}
{d\omega\over 2\pi}
{d\over d\omega}[\omega\coth({\omega\over 2T})]\nonumber\\
\sum_{\bf q}{Dq^2 V^R_2(\omega,{\bf q})\over 
[Dq^2-i\omega]^3}
={1\over 8\pi^2}\ln |\ln{\gamma\over T}|
\eea
which also shows a much weaker temperature dependence than 
the logarithmic weak-localization correction $\delta_{wl}\sigma_s(T)=
-\ln(\gamma\tau_{\phi}(T))/2\pi^2$  
controlled by the inelastic phase-breaking time $\tau_{\phi}(T)$ \cite{Senthil1}.

The latter 
can be estimated as a time interval for which the 
accumulated phase uncertainty 
\bea
\delta\Phi(t)=\int^{\infty}_{-\infty}{d\omega\over 2\pi}
\coth({\omega\over 2T})\nonumber\\
\sum_{\bf q}Im V^R_2(\omega,{\bf q})
{1-\exp(i\omega t-Dq^2t)\over [Dq^2-i\omega]^2}
\label{phase}
\eea
becomes of order unity \cite{Imry}. 
From the condition $\delta\Phi(\tau_\phi)\sim 1$, we obtain the estimate 
\be
\tau^{-1}_{\phi}(T)\sim {T\over 16\pi\sigma_s\ln^2\gamma/T}
\label{tauphi}
\ee
that should be contrasted 
against the apparent phase-breaking rate $\propto T^{1/3}$ 
found to describe both, experimentally and theoretically, 
quasiparticle transport in the normal state of the cuprates
\cite{Ong}.

Also, the dephasing time (\ref{tauphi}) differs from 
the inelastic energy relaxation  
(or, "out-scattering") time \cite{Blanter} 
\bea
\tau^{-1}_{\epsilon}=\int^{\infty}_{-\infty}{d\omega\over 2\pi}
[\coth({\omega\over 2T})-\tanh({\omega\over 2T})]\nonumber\\
\sum_{\bf q}Im V^R_2(\omega,{\bf q})
{1\over Dq^2-i\omega+\tau^{-1}_{\epsilon}}
\label{tau}
\eea
For $T<<\gamma/\sigma_s$
the solution of this self-consistent equation can be found 
in the form
\be
\tau^{-1}_{\epsilon}(T)= {T\ln\sigma_s\over 16\pi\sigma_s\ln^2\gamma/T}
\ee
Unlike $\tau_{\phi}$ which is determined by the processes with 
energy transters $\omega\sim
\tau^{-1}_{\phi}$,
$\tau_{\epsilon}$ receives contributions from 
the frequencies up to $\omega\sim T$.
In the ballistic ($T>\gamma$) regime, $\tau^{-1}_{\epsilon}(T)$
becomes a linear function of temperature, in agreement with  
the earlier theoretical 
\cite{Sachdev} and experimental \cite{Valla,Corson} results. 

In conclusion,
we carried out the analysis of the effects of
the experimentally relevant quasiparticle interactions  
mediated by fluctuations
of a secondary pairing order parameter ($is$) 
on the localization properties of 
the nodal quasiparticles in disordered high-$T_c$ superconducors. 
In the course of this study we revisited the problem of localization
of noninteracting random Dirac fermions and  
explicitly solved for all the relevant
diffusion and Cooperon modes in a 2D d-wave superconductor with arbitrarily
strong impurities. 
 We found that, despite being  
singular as a function of transferred energy and momentum, the 
interaction in question  
generates only subdominant, albeit positive, DOS, specific heat,
and conductivity corrections
and, therefore, does not necessarily alter the main predictions of the  
non-interacting localization scenario of Refs.\cite{Senthil1,Senthil2,Zirnbauer}.
Lastly, we provided the so far missing ingredient
of the localization theory by computing the inelastic 
dephasing and energy relaxation.

We note, in passing, that the latter does not account for the 
possibility of the formation of a strongly anisotropic 
network of delocalized resonant impurity states, as 
proposed in Ref.\cite{Balatsky}.
However, as far as the thermodynamic and transport properties 
of the rest of the electronic spectrum
are concerned, our findings provide further support for the 
attempts to find manifestations of the 
quasiparticle localization-related phenomena 
in the superconducting cuprates \cite{Bundschuh}.

Meanwhile, the high precision
heat transport measurements \cite{Taillefer,Aubin} might have already 
provided such an important evidence of weak localization
as positive magneto-(thermal)conductivity.
Should this interpretation of the data of Ref.\cite{Aubin} proposed in  
\cite{Bundschuh} prove to be correct, 
the phase relaxation time (\ref{tauphi}) may as well turn out 
to be experimentally accessible. 

The authors are grateful to A. Altland for his valuable comments
on the case of the $id$ secondary pairing and 
also on Eq.(11). This research was supported by the NSF under 
Grants DMR-0071362 (DVK) and DMR-9703388 (AGY), by RFBR (IVG), and partly by INTAS.

\wt
\end{document}